\documentclass[a4paper,numbers,final]{elsarticle}

\usepackage{graphicx}
\usepackage{amssymb}
\usepackage[english]{babel}
\usepackage[utf8x]{inputenc}
\usepackage{amsmath}
\usepackage{amsfonts}
\def\w{\omega}

\def\l{\lambda}
\def\bZ{{\bf Z}}
\def\d{{\rm \bf d}}

\newcommand{\blo}             [0]     {{\bf L_{0}}}
\newcommand{\bl}             [0]     {{\bf L}}
\newcommand{\bq}             [0]     {{\bf Q}}
\newcommand{\blh}             [0]     {{\bf Lh}}
\newcommand{\bqh}             [0]     {{\bf Qh}}
\def\f{{\bf f}}
\newcommand{\bjh}             [0]     {{\bf Jh}}
\newcommand{\bu}             [0]     {{\bf U}}

\newcommand{\br}             [0]     {{\bf R}}

\usepackage{listings}
\biboptions{square,numbers,compress,sort}
\journal{Journal of Sound and Vibration}
\begin{document}

\begin{frontmatter}

\title{A high-order, purely frequency based harmonic balance formulation for continuation of periodic solutions : the case of non polynomial nonlinearities}


\author[lma,amu]{Sami Karkar}
\ead{karkar@lma.cnrs-mrs.fr}
\author[lma,ecm]{Bruno Cochelin}
\ead{bruno.cochelin@ec-marseille.fr}
\author[lma]{Christophe Vergez}
\ead{vergez@lma.cnrs-mrs.fr}

\address[lma]{Laboratoire de M\'ecanique et d'Acoustique\\
CNRS -- UPR 7051\\
31 chemin J. Aiguier\\
13402 Marseille cedex 20\\
FRANCE}
\address[amu]{Aix-Marseille University, FRANCE}
\address[ecm]{\'Ecole Centrale Marseille\\
P\^ole de l'\'Etoile, Technop\^ole de Ch\^ateau-Gombert\\
13451 Marseille Cedex 20\\
FRANCE}

\begin{abstract}
In this paper, we extend the method proposed by Cochelin and Vergez in [A high order purely frequency based harmonic balance formulation for continuation of periodic solutions, {\it Journal of Sound and Vibration}, 324:243--262, 2009] to the case of non polynomial non-linearities. This extension allows for the computation of branches of periodic solutions of a broader class of nonlinear dynamical systems.

The principle remains to transform the original ODE system into an extended polynomial quadratic system for an easy application of the harmonic balance method (HBM). The transformation of non polynomial terms is based on the differentiation of state variables with respect to the time variable, shifting the nonlinear non polynomial non-linearity to a time-independent initial condition equation, not concerned with the HBM. The continuation of the resulting algebraic system is here performed by the Asymptotic Numerical Method (high order Taylor series representation of the solution branch) using a further differentiation of the non polynomial algebraic equation with respect to the path parameter.

A one d.o.f. vibro-impact system is used to illustrate how an exponential non-linearity is handled, showing that the method works at very high order, 1000 in that case. Various kinds of nonlinear functions are also treated, and finally the nonlinear free pendulum is addressed, showing that very accurate periodic solutions can be computed with the proposed method.
\end{abstract}


\end{frontmatter}

%\linenumbers


\section{Introduction\label{sec:intro}}
In a previous issue, Cochelin et Vergez presented \textit{A high order purely frequency-based harmonic balance formulation for continuation of periodic solutions} \cite{cochelin:2009b}, targeting nonlinear dynamical systems. The technique addresses autonomous or forced nonlinear dynamical systems, using an arbitrary high order Harmonic Balance Method (HBM, see \cite{urabe:1966,nakhla:1976}) together with the asymptotic numerical method (ANM, see \cite{cochelin:1994,cochelin:livre} for details) for the continuation.

The proposed method automatically derives the HBM set of algebraic equations to the desired order, provided that the governing equations are given in a quadratic formalism, ie, dynamical systems must be expressed as a set of first order nonlinear ordinary differential equations and nonlinear algebraic equations, all with purely quadratic polynomial nonlinearities at most (see eq. (3) in \cite{cochelin:2009b}).

The method has been used in various applications, such as the analysis of microbubbles in liquids by Pauzin et al. \cite{pauzin:2011}, or the nonlinear oscillations of nano- and micro-electromechanical devices  by Kacem et al. \cite{kacem:2011}. A purely frequency-based stability analysis was also proposed by Lazarus and Thomas \cite{lazarus:2010},  as a companion to the method presented in \cite{cochelin:2009b}.

\vskip\baselineskip
Whereas classical harmonic balance technique is often limited to very few harmonics, the proposed method has the advantage of allowing an arbitrary high order of the Fourier series. The difficulty is the need to recast the original system of equations (which is generally not quadratic), into a quadratic framework without changing the nature of the problem, ie, the recast should be only a change of formlution not a change of the original problem. The recast of polynomial, square root, and rational functions has been treated in the cited paper, defining intermediate variables and using quadratic polynomial algebraic equations. The authors also set up the basis for the recast of a sine nonlinearity, as encountered in the simple, free pendulum equation $\theta''(t) + \sin(\theta(t)) = 0$ (where the prime sign denotes time-derivation). They obtained a set composed of 4 nonlinear ordinary differential equations and 2 time-independent, nonlinear algebraic equations. But, the calculation was not carried out.

\vskip\baselineskip
In this paper, we propose a generalisation of the method proposed in \cite{cochelin:2009b} for the treatment of ODE with non polynomial nonlinearities. This generalisation greatly enlarges the field of application by overcoming the need for strictly quadratic reformulation. Therefore, the method can be used to compute periodic solutions families of most nonlinear dynamical systems. Notice that once a periodic solution is known, stability and bifurcation can be analyzed by using classical methods such as for instance, monitoring the eigenvalue of the monodromy matrix. This topic is not addressed hereafter.
%, as well as the stability of these solutions.

The present paper is organized as follows :  in section \ref{sec:expo}, the case of the exponential function is discussed as an introductory example and the periodic solutions of a one d.o.f vibro-impact system are calculated using up to 1000 harmonics (with MANLAB software) ; then, in section \ref{sec:general}, the method is generalized to a wider class of nonlinearities and in section \ref{sec:examples}, we show how this generalisation applies to the natural logarithm, trigonometric functions, as well as non-integer power functions ; and last, in section \ref{sec:pendul}, we use the proposed method to compute the periodic solutions of the simple free pendulum and we discuss the accuracy of the solution.

\section{An example using the exponential function: a regularised vibro-impact oscillator\label{sec:expo}}
We consider a one-degree-of-freedom, mass-spring oscillator which is limited to the half-plane $x<1$ by a rigid wall. The rigid wall reaction is modelled by an exponential function, with a coefficient $\alpha$ to tune the wall stiffness.

Let $x(t)$ denote the mass position, we look for periodic solutions of :
\begin{equation}
x''(t) = -x(t) -\l x'(t) - {\rm e}^{\alpha\left(x(t)-1\right)},
\label{eq:exp}\end{equation}
where $\l$ is a free parameter (see \ref{sec:vibroexp_annexe} for details on this equation).

\vspace{\parskip}
Before embarking in the treatment of (\ref{eq:exp}), we recall that two numerical methods are used in \cite{cochelin:2009b} : a high-order HBM technique and a high order Taylor series expansion for the continuation. The automation of these two techniques rely on the quadratic form of the nonlinearities. To comply with this framework, the method proposed by Cochelin et Vergez in \cite{cochelin:2009b} uses the following formalism:
\begin{equation}
\bf m(\bZ'(t)) = c_0 + \l c_1 + l_0(\bZ(t)) + \l l_1(\bZ(t)) + q(\bZ(t),\bZ(t))\label{eq:formalism}
\end{equation}
where $\bZ(t)$ is the unknown vector composed of time-dependent state variables, $\l$ is the continuation parameter, $\bf(c_0,c_1)$ are constant vectors, $\bf(l_0,l_1,m)$ are linear vector-valued operators, and $\bf q$ is a bilinear vector-valued operator.

\vskip\baselineskip
We will now present the successive recasts of system (\ref{eq:exp}) needed to obtain the required general form (\ref{eq:formalism}). 
It should be noticed in passing  that the quadratic form (\ref{eq:formalism}) has nothing to do with a second order Taylor expansion.
The passage form  (\ref{eq:exp}) to (\ref{eq:formalism}) should not introduce any approximation, it is just another formulation of the same original problem.
\subsection{First-order recast}
We introduce an additional variable $y(t)=x'(t)$ and rewrite equation (\ref{eq:exp}) as:
\begin{subequations}
\begin{align}
x'(t) =& y(t)\label{eq:exp-2}\\
y'(t) =& -x(t) -\l y(t) -{\rm e}^{\alpha\left(x(t)-1\right)}\label{eq:exp-2b}
\end{align}
\end{subequations}

\subsection{Quadratic recast of the exponential function}
Introducing the additional variable $e$ and its definition equation
\begin{equation}
    e(t)=\exp \Big[\alpha\big(x(t)-1\big)\Big]\label{eq:e1}
\end{equation}
we differentiate it with respect to the time variable to obtain
\[
    e'(t) = \alpha x'(t)  \exp \Big[\alpha\big(x(t)-1\big)\Big]
\]
or equivalently
\begin{equation}
    e'(t)=\alpha e(t) y(t)\label{eq:e2}
\end{equation}
Note that for equation (\ref{eq:e2}) to be exactly equivalent to (\ref{eq:e1}), the following initial condition must be added:
\begin{equation}
    e(0) = \exp \Big[\alpha\big(x(0)-1\big)\Big]\label{eq:e2-ci}
\end{equation}

We can now recast (\ref{eq:exp-2})-(\ref{eq:exp-2b}) into the following set of differential and algebraic equations:
\begin{subequations}
\begin{align}
x'(t) =& y(t)\label{eq:exp-3}\\
y'(t) =& -x(t) -\l y(t) -e(t)\label{eq:exp-3b}\\
e'(t) =& \alpha e(t) y(t)\label{eq:exp-3c}\\
e(0) =& {\rm e}^{\alpha\left(x(0)-1\right)}\label{eq:exp-3d}
\end{align}
\end{subequations}

At this point, the authors wish to underline the fact that the non polynomial nonlinearity has ``moved'' from an ODE equation (here: \ref{eq:exp-2b}) to an algebraic equation (here: \ref{eq:exp-3d}) which is not involved in the balance of harmonics: this is the key point of the method.

\subsection{Applying the harmonic balance method to the ODEs}
We will now apply the harmonic balance method (HBM) to the equations (\ref{eq:exp-3})-(\ref{eq:exp-3c}).

Denote $\bZ(t) = [x(t),y(t), e(t)]^T$, and the following operators:
\begin{align*}
{\bf m}(\bZ'(t)) = &[x'(t), y'(t), e'(t)]^T\\
{\bf c_0} = {\bf c_1} = &[0, 0, 0]^T\\
%c_1  = &[0, 0, 0]^T \\
{\bf l_0}(\bZ(t))  = &[y(t), -x(t)-e(t), 0]^T\\
{\bf l_1}(\bZ(t)) = &[0, -y(t), 0]^T \\
{\bf q}(\bZ(t),\bZ(t)) = &[0, 0, \alpha e(t) y(t)]^T
\end{align*}

The ODEs (\ref{eq:exp-3}-\ref{eq:exp-3c}) now take the required form:
\[
    \bf m(\bZ') = c_0 + \l c_1 + l_0(\bZ) + \l l_1(\bZ) + q(\bZ,\bZ).
\]

Eq. (\ref{eq:exp-3}-\ref{eq:exp-3b}) are treated classically, using the method proposed by Cochelin and Vergez in \cite{cochelin:2009b}, by balancing for the $H+1$ harmonics $0\leq h\leq H$ of $\bZ(t)$, where $H$ is the chosen truncation order of the Fourier series of the vector of (time-dependent) variables $\bZ(t)$:
\begin{equation}
\bZ(t) = \bZ_0 + \sum\limits_{h=1}^{H} \big( \bZ_{2h-1} \cos(k \w t) + \bZ_{2h} \sin(k\w t) \big)
\label{eq:serie_Fourier_Z}
\end{equation}

For equation (\ref{eq:exp-3c}), only harmonics $1\leq h \leq H$ need to be balanced. The reason is that the balance of the harmonic zero, which consists in equating the mean value of each side over a period, is not suitable here: both sides of the equation result from time-differentiation of a periodic quantity (namely $e(t)$), and has mean value which is therefore null. Thus, we only apply the HBM for harmonics $1\leq h \leq H$, where $H$ is the truncation order.

\vskip\baselineskip
The unknown (time-independent) variables are gathered in the state vector $\bu$:
\[
    \bu=\big[ \{\bZ_i\}_{i=0..2H}, \lambda, \omega\big],
\]
whose size is \mbox{$N_u$=$3(2H+1)+2$}.

The algebraic equations resulting from HBM application to  (\ref{eq:exp-3}-\ref{eq:exp-3b}-\ref{eq:exp-3c}) , are then automatically put into the standard ANM quadratic formalism:
\begin{equation}
\blo + \bl(\bu) + \bq(\bu,\bu) = 0\label{eq:man}
\end{equation}
with a constant vector $\blo$, a linear vectir-valued operator $\bl$, and a quadratic vector-valued operator $\bq$.

Now counting the number of algebraic equations:
\begin{itemize}
  \item $2(2H+1)$ for the full HBM applied to (\ref{eq:exp-3})-(\ref{eq:exp-3b})
  \item $2H$ for the partial HBM applied to (\ref{eq:exp-3c})
  \item $1$ for the time-independent equation (\ref{eq:exp-3d})
  \item $1$ for an additional phase equation (for instance: $y(0)=0$)
\end{itemize}
we reach a total number of $N_e$=$3(2H+1)+1$ equations.

Thus, we get $N_e=N_u-1$, as is needed for a 1D family of solutions.

\vskip\baselineskip
The reader shall notice that this algebraic system of $N_e$ equations for $N_u$ unknowns may now be solved by any continuation algorithm, such as for instance the very classical first order predictor with Newton corrector and  pseudo arc-length parametrization. 

In the following, we aim at using an ANM continuation method because of its robustness as compared to predictor-corrector algorithm. However, the algebraic system is not fully quadratic, ie, equation (\ref{eq:man}) is but not (\ref{eq:exp-3d}), hence, an additional step is necessary.

\subsection{Recast of the algebraic equation (\ref{eq:exp-3d})\label{sec:nl} for ANM continuation}
We will now address the last recast that concerns  non polynomial, nonlinear, \emph{algebraic} equation (\ref{eq:exp-3d}),  following the usual procedure presented in \cite{cochelin:livre} and \cite{manlab}. In the continuation process, the $N_u$ unknowns are function of a path parameter $a$. By differentiation of the non polynomial equation with respect to that path parameter $a$,  quadratic equation can be obtained as shown below.

For sake of clarity, let us define two new variables, $e_{t0}$ and $x_{t0}$, that represent $e(0)$ and $x(0)$ respectively:
\[
    e_{t0} = e_0 + \sum_{h=1}^H e_{2h-1} \qquad x_{t0} = x_0 + \sum_{h=1}^H x_{2h-1}.
\]
where $e_i$ (respectively $x_i$) denotes the $i$-th coefficient of the Fourier series of $e(t)$ (resp. $x(t)$) as defined for $\bZ$ in the equation (\ref{eq:serie_Fourier_Z}).

By differentiation, equation (\ref{eq:exp-3d}) is strictly equivalent to:
\begin{subequations}
\begin{align}
\forall a \in \mathbb{R}, \quad &{\rm d}e_{t0}(a) = \alpha \, e_{t0}(a) \, {\rm d}x_{t0}(a)\label{eq:diff1}\\
&e_{t0}(a\text{=}0) = {\rm e}^{\alpha(x_{t0}(a\text{=}0)-1)}\label{eq:diff2}
\end{align}
\end{subequations}
where ${\rm d}e= \frac{ d \, e}{d \, a}$ denote differentiation with rexpect to $a$.

Equation (\ref{eq:diff1}) is now quadratic in $\bf(U,dU)$ and is suitable for an easy and efficient computation of the Taylor series used for the continuation. The reader is referred to appendix \ref{sec:LhQh_annexe} for details on the formalism used to enter (\ref{eq:diff1}--\ref{eq:diff2}) in the MANLAB software.

\subsection{Periodic solutions of the regularised vibro-impact}
\begin{figure}[ht]
\begin{center}
    \begin{minipage}[c]{0.48\columnwidth}
      \includegraphics[width=\textwidth,clip=on]{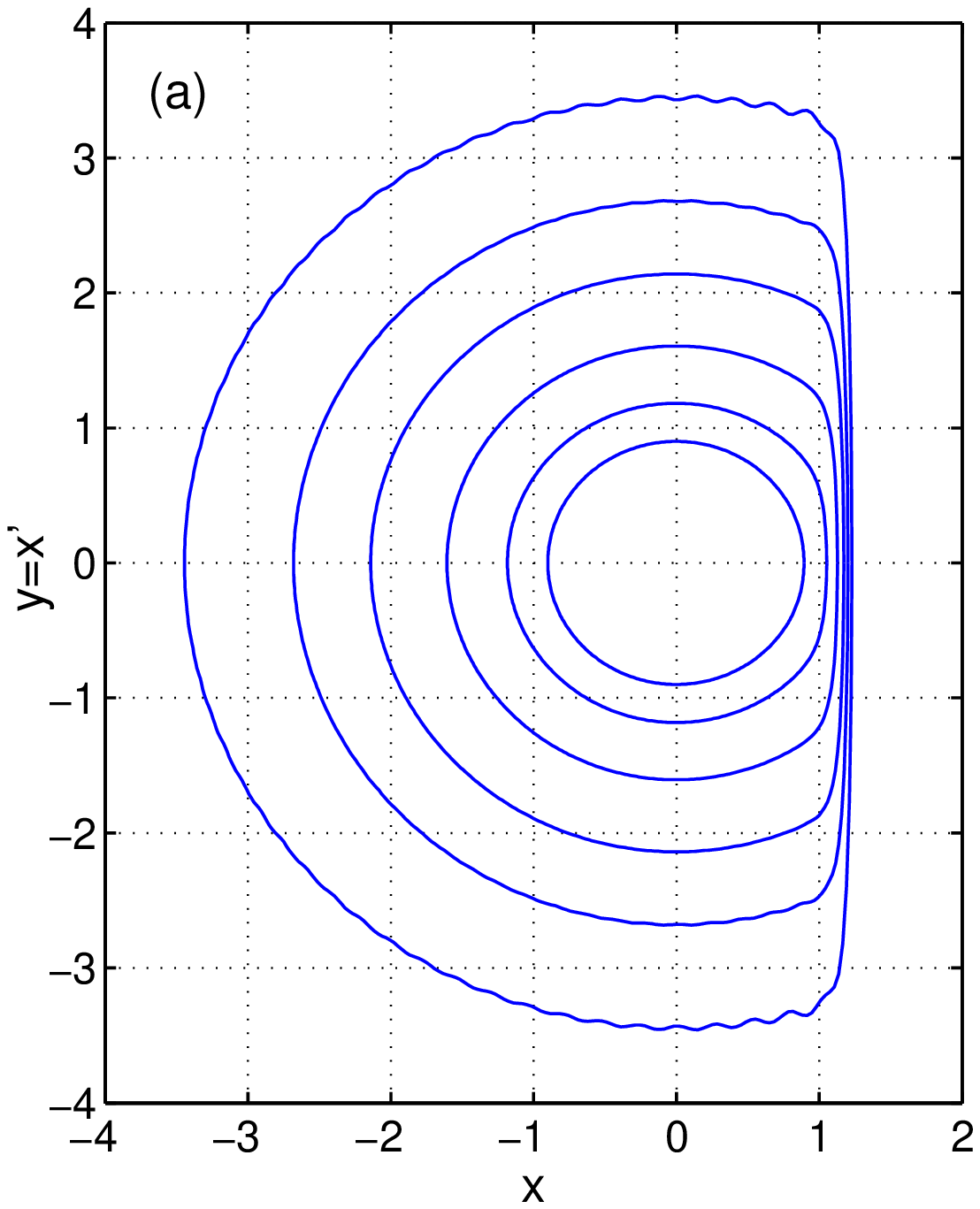}
    \end{minipage}
    \begin{minipage}[c]{0.48\columnwidth}
      \includegraphics[width=\textwidth,clip=on]{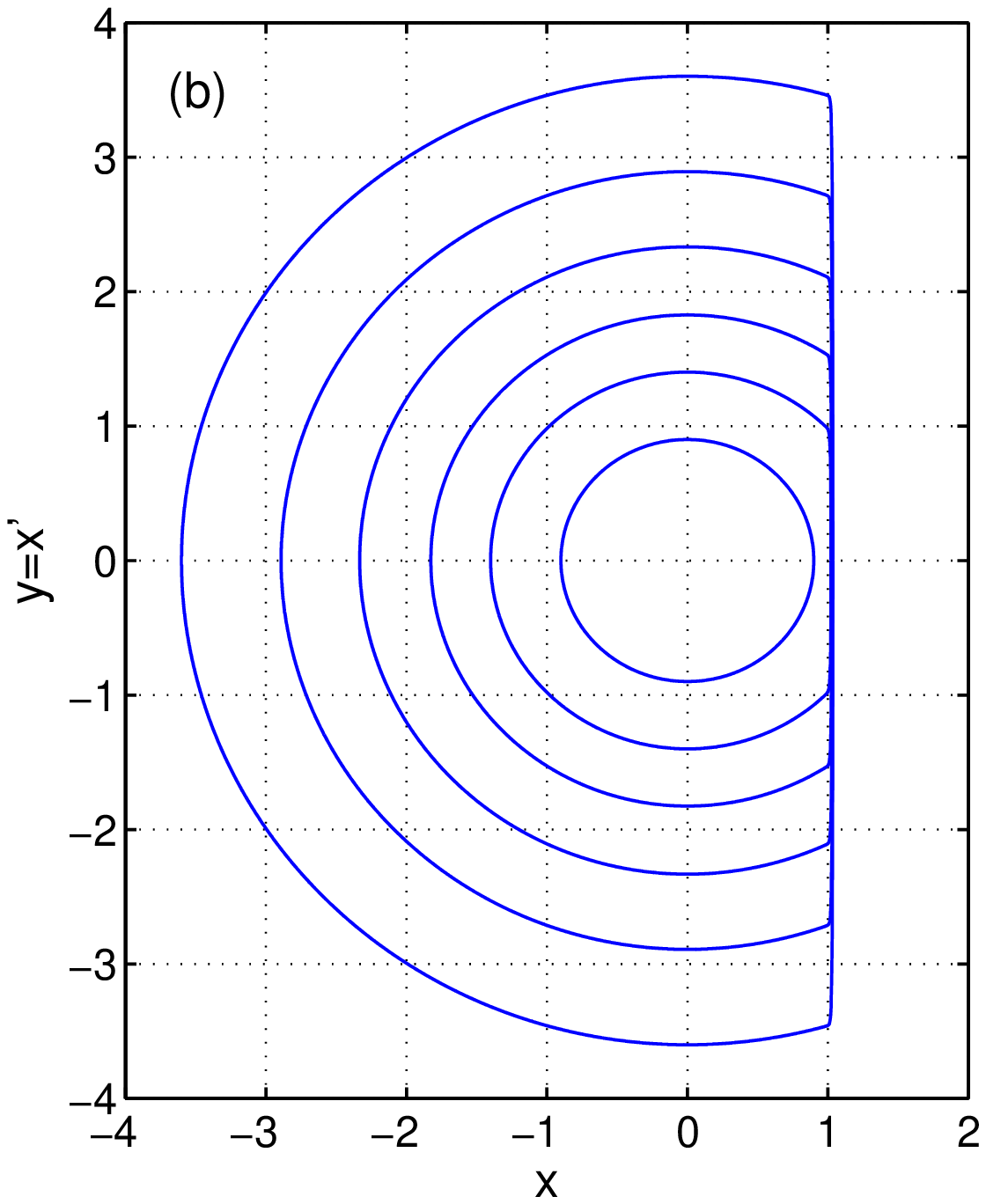}
    \end{minipage}
  \caption{\textbf{Phase portrait of the regularised vibro-impact:} family of periodic solutions in the phase plane $(u,v)$. (a): non stiff regularisation ($\alpha=20$) with $H=50$ harmonics. (b): stiff regularisation ($\alpha=200$) with $H=1000$ harmonics.}\label{fig:vibroexp}
\end{center}
\end{figure}
Figure \ref{fig:vibroexp} shows samples from the family of periodic solutions computed for two cases: a non stiff regularisation, with $\alpha=20$, using $H=50$ harmonics (a) ; and a stiff regularisation, with $\alpha=200$, using $H=1000$ harmonics (b). In both cases, the continuation was run with a threshold of $10^{-10}$ on the residue.

The phase portrait cycles are to be compared with those of a free, conservative non-regular vibro-impact system, i.e. a wall modelled with an impact law using a restitution coefficient equal to unity : the family of periodic solutions is composed of origin-centered circles, when the amplitude is less than unity, and origin centered arc of circles closed by a vertical segment along the line $x=1$, when the amplitude is higher than unity. 

%Results show that the non stiff regularisation captures well the general behaviour: circular orbits for $u<1$ and restricted move otherwise. But it fails to prevent the oscillator from penetrating the rigid wall: maximum penetration is almost 10\% of the $|u|_{max}$ amplitude.

%On the other side, the stiff regularisation, while demanding more harmonics, leads to a phase diagram in very good agreement with analytical results: penetration in the wall less than 1\% of the amplitude. Note that the same run with $H=400$ harmonics takes only a few minutes on a laptop computer and gives good results up to an amplitude of $|u|_{max}=3$.

\section{General treatment of nonlinear functions\label{sec:general}}
Here, we discuss the general method for the recast of most nonlinearities into quadratic formulation. Note that the quadratic recast of rational functions has been given by Cochelin and Vergez in \cite{cochelin:2009b}.

Let us consider a set of differential and algebraic equations:
\begin{equation}
F(u,v,g(u))=0 \label{eq:general-1}
\end{equation}
where \mbox{$v$=$u'$}, $F$ is at most quadratic in its arguments, and $g$ is any nonlinear function of $u$.

\subsection{First order derivative}
We add a new variable $w$ defined by \mbox{$w$=$g(u)$}. By time-derivation, we obtain:
\begin{subequations}
\begin{align}
w' &=\frac{\partial g}{\partial u}(u)v\label{eq:general-2}\\
w(0) &= g(u(0))\label{eq:general-2ci}
\end{align}
\end{subequations}

If $\partial g/\partial u$ can be written as a rational function of ($u,v,w$), then the equation \mbox{$x=\partial g/\partial u$} can be recast into quadratic equations (possibly using additional variables). Consequently, the time-dependent equation (\ref{eq:general-2}) becomes $w'=xv$, which is quadratic in $x$ and $v$.

We apply the HBM to this equation, but only for harmonics $1\leq h\leq H$, while the mean value (harmonic zero) will be constrained by the initial condition (\ref{eq:general-2ci}). The equation count is then equal to $2H+1$, as in a standard HBM applied to a unique equation, which matches the number of variables : the $2H+1$ coefficients of the Fourier series of $w(t)$ up to harmonic $H$. In the case of autonomous systems, the period (or equivalently the angular frequency) is also unknown and one need to add a phase equation, as explained by Doedel in \cite{doedel:2010} as well as Cochelin and Vergez in \cite{cochelin:2009b}.

If additional variables were to be used for the quadratic recast of $x$, all corresponding algebraic equations will be treated with full HBM, including the balance of the harmonic zero (the mean values).

\subsection{Second order derivative}
If $x=\partial g/\partial u$ cannot be expressed as a quadratic polynomial of the current variables, we differentiate it with respect to $t$:
\begin{subequations}
\begin{align}
x' &=\frac{\partial^2 g}{\partial u^2}(u)v\label{eq:general-3}\\
x(0) &= \frac{\partial g}{\partial u}(u(0))\label{eq:general-3b}
\end{align}
\end{subequations}

If $\partial^2g/\partial u^2$ can be expressed as a rational function of \mbox{($u,v,w,x$)}, then we define \mbox{$y$=$\partial^2 g/\partial u^2$} and the previous results apply: the time-dependent equation in (\ref{eq:general-3}) is quadratic in $y$ and $v$.

We then have two differential equations, namely \mbox{$w'$=$xv$} and \mbox{$x'$=$yv$}, with two associated initial conditions. The differential equations are treated using HBM for harmonics 1 and higher, while the initial conditions will constrain the mean values of $w$ and $x$ (the nonpolynomial, nonlinear, algebraic equation is addressed by differentiation, as explained in section \ref{sec:nl}).

If additional variables were used for the recast of the rational function $y$, all corresponding algebraic equations will be treated with full HBM, including the balance of the harmonic 0 (the mean values).

\section{Recast of a few common non-polynomial nonlinearities\label{sec:examples}}
For the quadratic recast of the exponential function, the reader is referred to section \ref{sec:expo}.

\subsection{Natural logarithm}
Given $w$ defined as $w(t)=\mathrm{ln}(u(t))$ (assuming $u>0$). For sake of simplicity, we do not write explicitly the time dependence in the following. Differentiating with respect to the time variable, the definition equation becomes $w'=u'/u$, or equivalently, using $x=w'$:
\[
    \left\{
\begin{array}{l}
    u'=xu\\
    w(t=0) = \mathrm{ln}(u(t=0))
\end{array}
    \right.
\]
which is quadratic in $u$ and $x$.

\subsection{Non-integer power}
Given $w(t) = u(t)^\alpha$ where $\alpha\in \mathbb{R}$ is a constant, then $w'=\alpha u^{\alpha-1}u'$. Using $v=u'$ and $x=w'$, one gets:
\[
    \left\{
\begin{array}{l}
    ux=\alpha w v\\
    w(t=0) = u(t=0)^\alpha
\end{array}
    \right. 
\]
which is quadratic in ($u,v,w,x$).

\subsection{Trigonometric functions}
Given $s(t)=\sin(u(t))$ and $c(t)=\cos(u(t))$, we introduce $v(t)=u'(t)$ and time-derivation of the definition equations of $s$ and $c$ gives:
\[
    \left\{
\begin{array}{l}
    s' = c v\\
    c' = -s v\\
    s(t=0) = \sin(u(t=0))\\
    c(t=0) = \cos(u((t=0))
\end{array}
    \right. 
\]
which are obviously quadratic in ($c,v$) and ($s,v$) respectively.

As for $w(t)=\tan(u(t))$, time-differentiation leads to $w'=(1+w^2)v$. Using an additional variable $x=1+w^2$, one gets the quadratic equation:
\[
    \left\{
\begin{array}{l}
    w'=xv\\
    w(t=0) = \tan(u(t=0)).
\end{array}
    \right. 
\]

\section{Periodic solutions of the simple, free, nonlinear pendulum\label{sec:pendul}}
Denoting $\theta$ the angle between the current position and the lower, rest position, the equation of motion of the free pendulum writes:
\begin{equation}
\theta''(t) + \sin{\big[\theta(t)\big]} = 0 \label{eq:pend1}
\end{equation}

Adding a damping parameter $\l$, that will vanish along the family of periodic solutions, as explained in section \ref{sec:vibroexp_annexe}, the equation of motion becomes:
\begin{equation}
\theta''(t) + \l \theta'(t) + \sin{\big[\theta(t)\big]} = 0 \label{eq:pend2}
\end{equation}

Using three additional variables, \mbox{$v(t)$=$\theta'(t)$}, \mbox{$s(t)$=$\sin{\big[\theta(t)\big]}$} and \mbox{$c(t)$=$\cos{\big[\theta(t)\big]}$}, we recast the equation (\ref{eq:pend2}) into the following quadratic, differential-algebraic system:
\begin{subequations}
\begin{align}
\theta'(t) &= v(t) \label{eq:pendquad-a}\\
v'(t) &= -s(t) - \l v(t)\label{eq:pendquad-b}\\
s'(t) &= c(t) v(t)  \label{eq:pendquad-c}\\
c'(t) &= -s(t) v(t)  \label{eq:pendquad-d}\\
s(0) &= \sin{\big(\theta(0)\big)}  \label{eq:pendquad-e}\\
c(0) &= \cos{\big(\theta(0)\big)}  \label{eq:pendquad-f}
\end{align}
\end{subequations}

Using the proposed method, we apply:
\begin{itemize}
  \item the full HBM to equations (\ref{eq:pendquad-a}-\ref{eq:pendquad-b})
  \item the balance of the harmonics 1 and higher to equations (\ref{eq:pendquad-c}-\ref{eq:pendquad-d})
\end{itemize}
and since the frequency is also unknown, we add a phase equation:
\begin{equation}
v(0) = 0.
\end{equation}

Finally, the initial conditions (\ref{eq:pendquad-e}-\ref{eq:pendquad-f}) are differentiated with respect to the path parameter. The equations are then in the right (quadratic) form for applying the ANM continuation, ie, computing a high orde Taylor series of the unknowns with respect to the path parameter:
\[
  \left\{
\begin{array}{l}\bf
  \blo + \bl(\bu) + \bq(\bu,\bu) = {\bf 0}\\
  \blh(\d\bu) = \bqh(\bu,\d\bu)
\end{array}
\right.
\]

The total number of equation is then:
\[
\begin{array}{ll}
    N_e &= \underbrace{\underbrace{2(2H+1)}_\text{full HBM} + \underbrace{2(2H)}_{\text{HBM }h>0}
             + \underbrace{\quad 1 \quad}_\text{phase eq.}}_\text{in L0/L/Q} + \underbrace{\quad 2 \quad}_\text{in Lh/Qh}\\
        &= 4(2H+1)+1
\end{array}
\]
while the number of variables is 
\[
\begin{array}{ll}
    N_u &= \underbrace{\quad\qquad 4(2H+1) \qquad\quad}_{\text{Fourier components of } \theta,v,s,c} + \underbrace{1}_{\l} + \underbrace{1}_{\w}\\
        &= 4(2H+1)+2
\end{array}
\]

\begin{figure}[ht]
\begin{center}
    \begin{minipage}[c]{0.48\columnwidth}
      \includegraphics[width=\textwidth,clip=on]{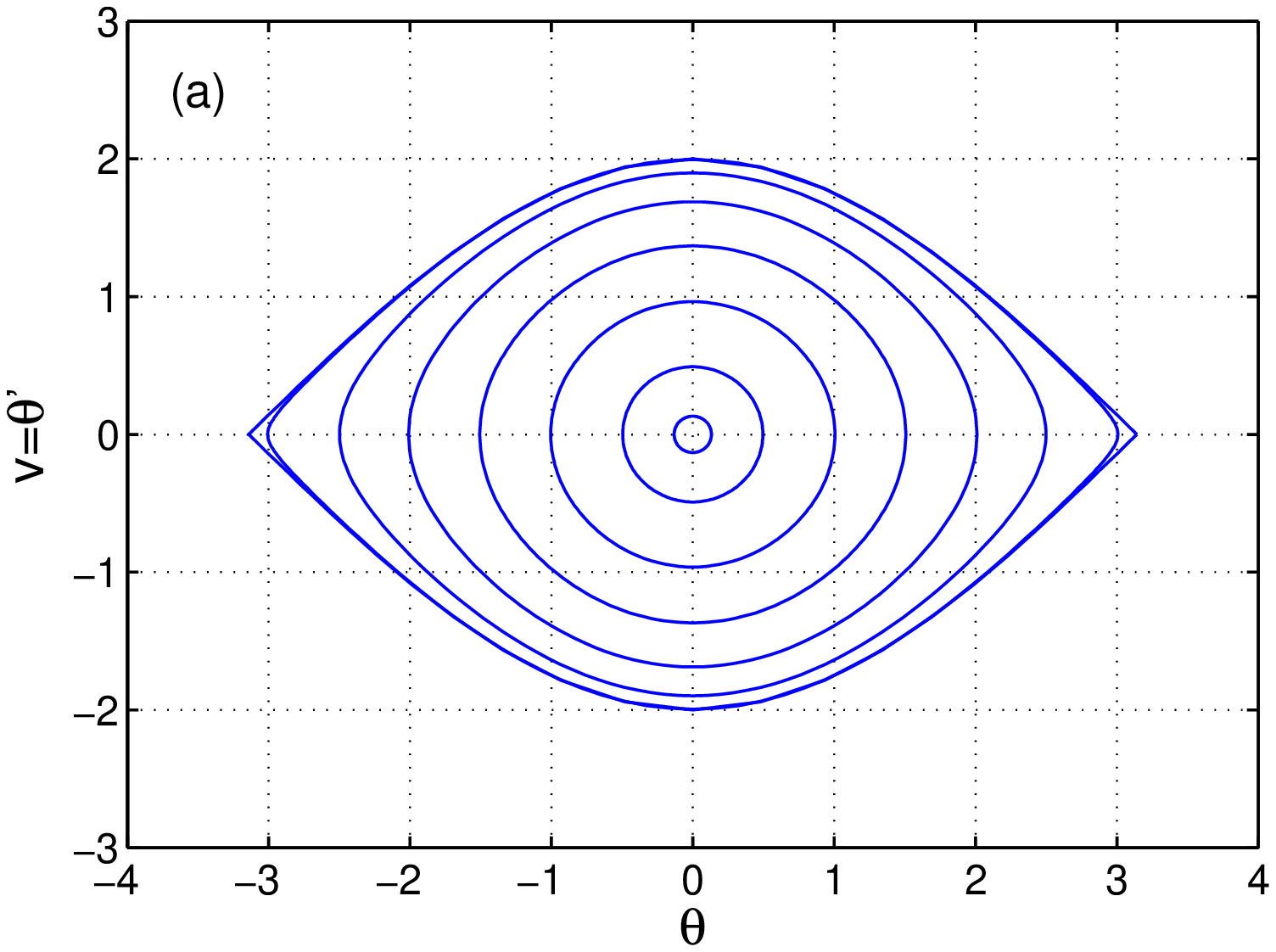}
    \end{minipage}
    \begin{minipage}[c]{0.48\columnwidth}
      \includegraphics[width=\textwidth,clip=on]{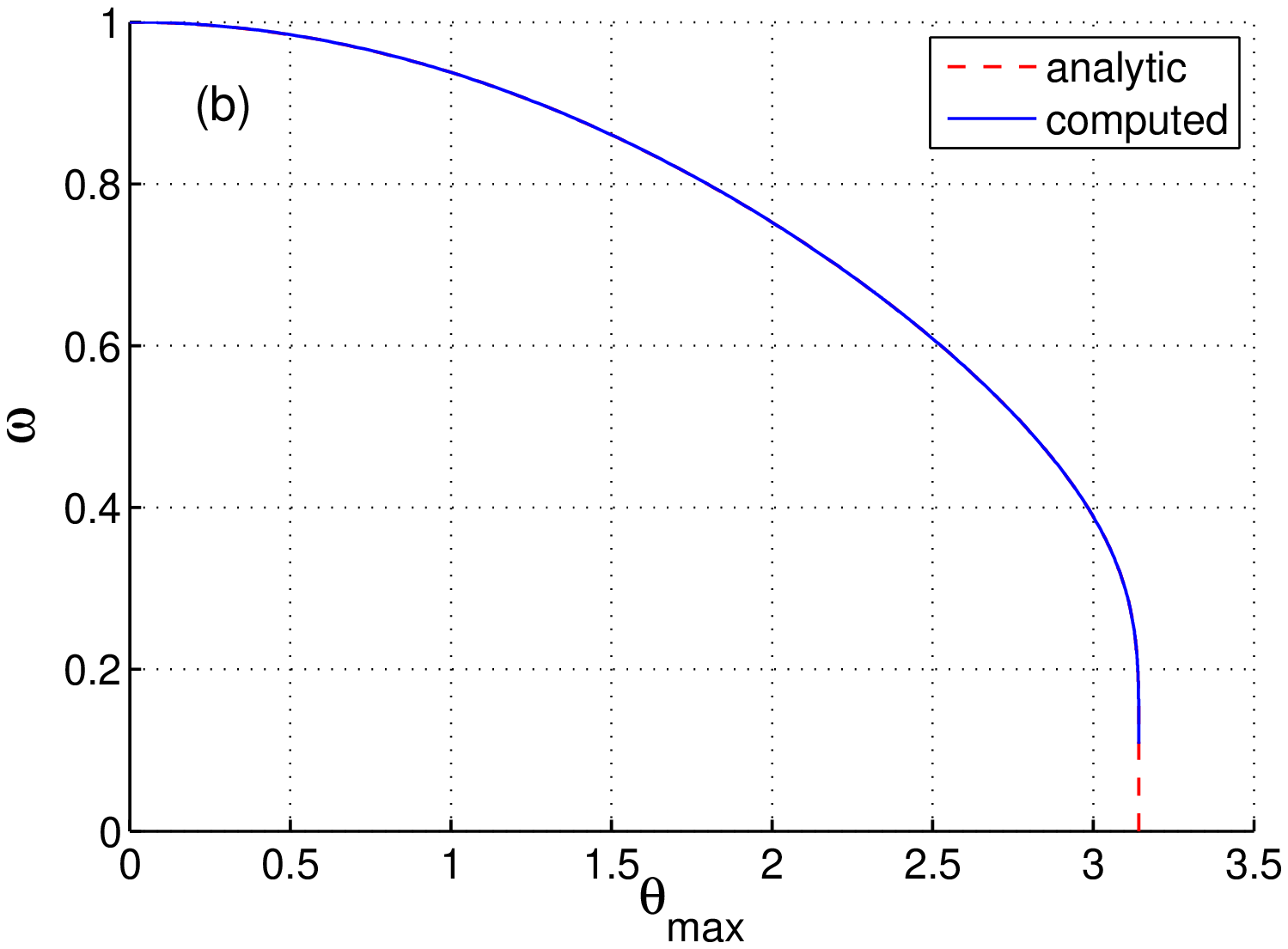}
    \end{minipage}
  \caption{\textbf{Periodic solutions of the nonlinear free pendulum.} (a): phase portrait $(\theta,v)$ showing a few samples of the periodic solution family. (b): amplitude-frequency diagram, the computed curve is superimposed on that of the analytic formula. HBM using $H=100$ harmonics, ANM-threshold: $10^{-15}$, the residue $||\br||$ was kept under $10^{-14}$.}\label{fig:pendule}
\end{center}
\end{figure}
Figure \ref{fig:pendule} shows the phase portrait and the amplitude-frequency diagram of the periodic solutions family of the nonlinear, free pendulum system, computed with $H=100$ harmonics. The theoretical amplitude-frequency diagram of this system (dashes) corresponds to the following analytic formula:
\[
    T(\theta_{max}) = \frac{2}{\pi} {\bf K}\Big(\sin^2(\theta_{max}/2)\Big),
\]
where ${\bf K}$ is the complete elliptic integral of the first kind (see \cite{belendez:2007} for instance). The computed curve (plaine line) is superimposed on the theoretical one.

%, and for each control point a precise (but expansive) time-integration was carried out, resulting in a reference solution. The relative error on the period $\epsilon_T=|(T_{ref}-T)/T_{ref}|$ was evaluated, as well as the relative error on the solution $\epsilon_{\theta}=||\theta_{ref}(t)-\theta(t)||/||\theta_{ref}(t)||$ (where $||\cdot||$ is the euclidean norm of the vector of the values of $\cdot$ sampled on $2H$ equally spaced points over a period).
\vskip\baselineskip
The branch was computed with $29$ step of continuation. The computation was stopped when the relative error between the computed angular frequency and its theoretical value (given the amplitude) reached $0.1\%$, which occurs at the solution point:
\[
    \begin{pmatrix}\theta_{max}\\\w\end{pmatrix}=\begin{pmatrix}0.999998\pi\text{ rad}\\0.112801\text{ rad/s}\end{pmatrix}
\]
where the norm of the residual vector is $||\br||$=$2.76$ $10^{-15}$. Increasing the number of harmonics only leads to a closer approach to the limit point $(\theta_{max},\w)$=$(\pi,0)$, where the period is infinite\footnote{This corresponds to a heteroclinic connection between the two saddle-nodes \mbox{$(\theta_{SN},v_{SN})\in\{(-\pi,0),(\pi,0)\}$}}.

 Notice that two additional variables must be introduced for this simple one d.o.f examle. For a multiple d.o.f. system with many kinds of non-polynomial nonlinearities, the quadratic recast has to be applied to each individual nonlinear term. This may require a great number of additional variables and the transformed system may be much larger than the original one. This is however the price to pay with this method, the more complex the original system, the larger the transformed  quadratic system.

\section{Conclusion}
In this paper, we extended the works of Cochelin and Vergez presented in \cite{cochelin:2009b} to the case of general nonlinearities.

A vibro-impact with exponential regularization was presented, and its periodic solutions were computed with the proposed generalization. The case of a very stiff regularization demonstrated the capabilities of this numerical tool to deal with a very high number of harmonics in the harmonic balance method.

Finally, we showed how to apply the method for the continuation of the periodic solutions of the simple, free, nonlinear pendulum. Our results confirm that the harmonic balance method with a high number of harmonics is both affordable and well suited to the ANM continuation framework. It has been successfully implemented in the MANLAB software which provide an interactive graphical user interface for the continuation, in a widely used programming environment.

The principal limitation of the proposed method is due to the need of an algebraic relation between the nonlinear function and its derivative, or its primitives and the state variables. However, the numerous examples treated in this paper show a variety of such functions that might appear in nonlinear dynamical systems.

Concerning future work, the companion frequency-based stability analysis presented in \cite{lazarus:2010} could be adapted to general nonlinearities, following the same approach. Also, automatic differentiation, such as the DIAMANT package proposed by Charpentier et al in \cite{charpentier:2008}, could give a mean to simplify the user input to the bare nonlinear functions, letting the user free of any recast.

%\section*{Acknowledgements}
%The author wish to thank .... for the and fruitful discussions.
\clearpage
\appendix
\renewcommand\thesection{\Alph{section}}
\makeatletter
\def\@seccntformat#1{\csname Pref@#1\endcsname \csname the#1\endcsname\quad}
\def\Pref@section{Appendix~}
\makeatother

\section{Vibro-impact with exponential wall reaction\label{sec:vibroexp_annexe}}
\subsection{Model}
We consider a one-degree-of-freedom, mass-spring oscillator which is limited to the half-plane $x<1$ by a rigid wall, where $x(t)$ denotes the position of the mass.

The rigid wall reaction is modelled by an exponential function, with a coefficient $\alpha$ to tune the wall stiffness :
\[
    F_{w}(x) = -{\rm e}^{\alpha(x-1)},
\]

Thus, the regularised vibro-impact system is governed by the following equation:
\begin{equation}
x''(t) = -x(t) - {\rm e}^{\alpha\left(x(t)-1\right)},
\label{eq:exp-1}\end{equation}
where the prime sign denotes time-differentiation. The force $F_w(x)$ that reflects the wall effect derives from a potential energy so that problem (\ref{eq:exp-1}) keeps the property of being energy-conservative.

\subsection{Dissipative recast for continuation}
Mu\~noz-Almaraz et al. \cite{munoz:2003} showed that, in conservative Hamiltonian systems, periodic orbits generally belong to a one-dimensional family of periodic solutions, parametrised by the value of the first integral (here, the total energy), which is not an explicit parameter of the system. To compute this family of periodic solutions in the standard continuation framework $R(x(t),\l)=0$, we perturb the initial equation with a damping term added to the right-hand side of (\ref{eq:exp-1}). The system is then embedded into a general, dissipative system:
\begin{equation}
x''(t) = -x(t) -\l x'(t) - {\rm e}^{\alpha\left(x(t)-1\right)},
\label{eq:exp-1bis}\end{equation}
where $\l$ is a free parameter of the continuation.

The perturbed system (\ref{eq:exp-1bis}) possesses periodic solutions that are exactly those of the unperturbed, conservative system (\ref{eq:exp-1}), if and only if $\l=0$. This way, the additional parameter $\l$ allows us to compute the periodic solutions of the conservative system (\ref{eq:exp-1}) using the classical framework for dissipative systems possessing an explicit control parameter.

\section{Classical ANM: quadratic framework\label{sec:MAN_annexe}}
The reader is referred to Cochelin and Vergez \cite{cochelin:2009b} (sections 2.4 and 2.5, pp.248-250), for the details concerning the principle and the implementation of the ANM in the classical, quadratic framework.

\section{Extended ANM framework\label{sec:LhQh_annexe}}

%Then, given a set of $N_e'$ nonlinear equations $f(\bu)=0$, we may differentiate it w.r.t. $a$:
%\[ \frac{df}{da} = 0 \qquad \text{and} \quad f(\bu(a=0))=0 \] or equivalently:
%\begin{subequations}
%\begin{align}
%\frac{df}{d\bu} \frac{d\bu}{da} = 0 \label{eq:f_diff_a}\\
%f(\bu(a=0)) = 0 \label{eq:f_diff_b}
%\end{align}
%\end{subequations}

%Note that $\frac{df}{d\bu} : \mathbb{R}^{N_e+1} \rightarrow \mathbb{R}^{N_e'}\mathbb{R}^{N_e+1}$ is a vector valued function of $\bu$. We assume the strong hypothesis that $df/d\bu$ is \emph{linear} in $\bu$. Then equation (\ref{eq:f_diff_a}) is a bilinear, vector valued operator of $(\bu,d\bu/da)$.

\subsection{Principle of the series computation}
Given a nonlinear system $\f(\bu)=0$ with $N_d$ equations and $N_u$ unknowns, whose differentiated form reads:
\begin{equation}
\blh(\d\bu) = \bqh(\bu,\d\bu) \label{eq:ext_man}
\end{equation}
where $\blh : \mathbb{R}^{N_u} \rightarrow \mathbb{R}^{N_d}$ is a linear, vector-valued operator ; $\bqh : \mathbb{R}^{N_u} \times \mathbb{R}^{N_u} \rightarrow \mathbb{R}^{N_d}$ is a bilinear, vector-valued operator.

Assuming a known regular solution $\bu_0$ of this system, we write the branch of solutions passing through this point as a Taylor series expansion:
\begin{equation}
\bu(a) = \bu_0 + a \bu_1 + a^2 \bu_2 + a^3 \bu_3 + \dots + a^n \bu_n \label{eq:U_serie}.
\end{equation}
where the branch is parametrised using the pseudo-arclength parameter $a$ defined as:
\begin{equation}
    a=(\bu-\bu_0)^t \bu_1.
    \label{eq:pseudo-arclength}
\end{equation}

Differentiating $\bu$ reads:
\begin{equation}
\d\bu = \{\bu_1 + 2 a \bu_2 + 3 a^2 \bu_3 + ... + n a^{n-1} \bu_n\} da. \label{eq:dU_serie}
\end{equation}

Then, substituting both (\ref{eq:U_serie}) and (\ref{eq:dU_serie}) in system (\ref{eq:ext_man}) and equating each power of $a$ (up to order $n$) to zero gives:
\begin{itemize}
  \item{power $0$:} $\blh(\bu_1) - \bqh(\bu_0,\bu_1) = 0 $, which can also be written \mbox{$\bjh_{\bu_0}.\bu_1 = 0$} where $\bjh_{\bu_0} \in \mathbb{R}^{N_d \times N_d+1}$ is the jacobian matrix of $f$ evaluated at $\bu_0$. This linear equation in $\bu_1$ thus gives the term of \emph{order $1$} of (\ref{eq:U_serie}).
  \item{power $1 \leq p \leq n-1$:} $\bjh_{\bu_0}.\bu_{p+1} = \Sigma_{i=1}^{p}\frac{p+1-i}{p+1}\bqh(\bu_i,\bu_{p+1-i})$. This linear equation gives the term of \emph{order $p+1$} of (\ref{eq:U_serie}).
\end{itemize}
The original nonlinear problem is thus replaced by a cascade of $n$ linear systems, which all share the same matrix: $\bjh_{\bu_0}$.

However, at each order, the linear systems have $N_d+1$ unknowns and only $N_d$ equations. For each linear system, the additional equation is obtained by substituting the series (\ref{eq:U_serie}) into the definition (\ref{eq:pseudo-arclength}) of the path parameter $a$:
\begin{itemize}
\item {\bf order 1 :} $\bu_1^t  \bu_1 = 1$
\item {\bf order $\bf 2 \leq p \leq n$ :} $\bu_1^t \bu_p = 0$
\end{itemize}

\subsection{Implementation in MANLAB: the example of the vibro-impact}
As for the classical framework, the only user input to the MANLAB software consists in M-functions for the operators $\blh$ and $\bqh$, as well as for $f$ (for the residue computation only), and a starting point $\bu_0$.

In the case of the vibro-impact system presented in section \ref{sec:expo}, some equations are quadratic (those resulting from the HBM and those for the definition of $e_{t0}$ and $x_{t0}$) while the last one is not. We thus separate the equations in two parts: those resulting from the HBM, that will appear in the $\bl0$, $\bl$ and $\bq$ operators, and the last one, that will appear in the $\blh$, $\bqh$ operators.

For the present problem, the state vector is:
\[
    \begin{array}{l}
    \bu = (\underbrace{x_0,y_0,e_0}_{\bZ_0},\underbrace{x_1,y_1,e_1}_{\bZ_1},\underbrace{x_2,y_2,e_2}_{\bZ_2}, \cdots, \qquad \qquad \qquad  \\
     \qquad \qquad \qquad \underbrace{x_{2H-1},y_{2H-1},e_{2H-1}}_{\bZ_{2H-1}},\underbrace{x_{2H},y_{2H},e_{2H}}_{\bZ_{2H}}, \l, \w, e_{t0}, x_{t0} )
    \end{array}
\]
and its size is $N_u = 3(2H+1)+4$.

For the quadratic part of the problem, the subsystem contains the following number of equations:
\[
N_e=\underbrace{2(2H+1)}_\text{full HBM} + \underbrace{\quad 2H \quad}_{\text{HBM }h\neq0} + \underbrace{\qquad 2\qquad }_{\text{def. of } e_{t0},x_{t0}} + \underbrace{\quad 1 \quad}_\text{phase eq.} = 3(2H+1)+2.
\]
The content of \verb!L0.m!, \verb!L.m! and \verb!Q.m! are not listed here, as it is the direct result of the harmonic balance as presented in \cite{cochelin:2009b}. These three functions return a vector of size $N_e-1$.

\vskip\baselineskip
For the non quadractic part, the subsystem contains $N_d=1$ equation. The content of {\tt Lh.m}, {\tt Qh.m} and {\tt f.m} is listed below, with $\alpha=200$:

\begin{verbatim}
    function [Lh] = Lh(dU)
    Lh=zeros(1,1);
    Lh = dU(end-1);
    
    
    function [Qh] = Qh(U,dU)
    Qh = zeros(1,1);
    Qh = 200*U(end-1)*dU(end);
    
    
    function [f]  = f(U)
    f = zeros(1,1);
    f = exp(200*(U(end)-1));
\end{verbatim}

\clearpage
\bibliography{exp-nl}
\clearpage
\listoffigures
\end{document}